\documentclass[submission, Phys]{SciPost}

\usepackage[utf8]{inputenc}

\usepackage{amsmath,amssymb,amsfonts,amscd,dsfont,bm}     % Standard mathematics 
\allowdisplaybreaks     %Allows pagebreak during multiline math enviroments
\usepackage{mathrsfs}
\usepackage{hyperref}
\usepackage{slashed}
\usepackage{youngtab}
\usepackage{physics}
\usepackage{tensor}

\usepackage{graphicx}
\graphicspath{{./Figures/}}

\usepackage[small,nohug,heads=littlevee]{diagrams}
\diagramstyle[labelstyle=\scriptstyle]

\begin{document}

\begin{center}{\Large \textbf{
Invariant Renormalization-Group improvement
}}\end{center}

\begin{center}
Aaron Held*
\end{center}

\begin{center}
Theoretical Physics, Blackett Laboratory,
\\
Imperial College London, SW7 2AZ London, U.K.
\\
* a.held@imperial.ac.uk
\end{center}

\begin{center}
\today
\end{center}

\section*{Abstract}
{\bf
Renormalization-Group (RG) improvement has been frequently applied to capture the effect of quantum corrections on cosmological and black-hole spacetimes. This work utilizes an algebraically complete set of curvature invariants to establish that: On the one hand, RG improvement at the level of the metric is coordinate-dependent. On the other hand, a newly proposed RG improvement at the level of curvature invariants is coordinate-independent. Spherically-symmetric and axially-symmetric black-hole spacetimes serve as physically relevant examples.
}

\vspace{10pt}
\noindent\rule{\textwidth}{1pt}
\tableofcontents\thispagestyle{fancy}
\noindent\rule{\textwidth}{1pt}
\vspace{10pt}

\section{Introduction}

In view of the Hawking-Penrose singularity theorems~\cite{Penrose:1964wq, Hawking:1965mf, Geroch:1966ur, Hawking:1969sw}, the occurrence of singularities in General Relativity (GR) has been established with mathematical rigor. These theorems rely on the central assumption that the gravitational force is universally attractive, i.e., always leads to a focusing of adjacent lightlike geodesics.

As singularities are unphysical, they have to be resolved by the onset of some form of new physics. The above assumption of a universally attractive gravitational force hints at a resolution: If the growth of some physical quantity $\mathcal{K}$, which diverges at the GR singularity, is associated with a weakening of the effective gravitational force, this may mitigate or even remove the singularity in the respective beyond-GR theory. This effect might then be captured by a $\mathcal{K}$-dependent improved Newton coupling $G(\mathcal{K})$, cf.~\cite{Markov:1985py} for an early account, and~\cite{Bardeen:1968,Dymnikova:1992ux,Hayward:2005gi,Simpson:2019mud} for ad-hoc proposals for regular black-hole spacetimes. 

Noticing that the sign of quantum-corrections indeed indicates a weakening of the gravitational force~\cite{Gastmans:1977ad, Christensen:1978sc, Donoghue:1993eb, Reuter:1996cp, Bosma:2019aiu}, the groundbreaking work of Bonanno and Reuter~\cite{Bonanno:1998ye, Bonanno:2000ep} initiated the use of Renormalization-Group (RG) improvement\footnote{In non-gravitational massless $\phi^4$-theory, Coleman and Weinberg have formally shown~\cite{Coleman:1973jx} that replacing the theories coupling constant by its renormalization-scale dependent equivalent amounts to a re-summation of large logarithms. The resulting RG-improved effective potential remains valid as long as the respective running coupling remains within perturbative control. Similarly, RG-improvement recovers the quantum corrections to the Coulomb potential of massless QED~\cite{Dittrich:2014eff}.} in the context of GR to model the effect of said repulsive quantum fluctuations on classical geometries. 

RG-improvement of classical spacetimes has since been applied to the Schwarzschild spacetime~\cite{Bonanno:1998ye, Bonanno:2000ep,Platania:2019kyx}, to Schwarzschild-(A)dS black holes~\cite{Koch:2013owa, Kofinas:2015sna, Torres:2017ygl, Pawlowski:2018swz, Adeifeoba:2018ydh} to the Vaiyda metric~\cite{Bonanno:2006eu}, to spherically-symmetric gravitational collapse~\cite{Torres:2014gta,Torres:2014pea,Torres:2015aga,Bonanno:2016dyv,Bonanno:2017zen}, to rotating Kerr black holes~\cite{Reuter:2006rg, Reuter:2010xb, Falls:2012nd, Pawlowski:2018swz, Held:2019xde}, to higher-dimensional black holes~\cite{Falls:2010he,Litim:2013gga}, to stellar interiors \cite{Fayos:2011zza,Bonanno:2019ilz} and in the context of cosmology \cite{Bonanno:2001xi,Bonanno:2001hi,Bonanno:2002zb,Reuter:2003ca,Reuter:2004nv,Babic:2004ev,Bonanno:2012jy,Bonanno:2015fga}, cf.~\cite{Platania:2020lqb} for a recent review. In brief, RG-improvement consists of two central physical inputs, i.e.,
\begin{itemize}
	\item
	\textbf{(I) scale dependence:} where, for instance, the classical Newton coupling $G_0$ is replaced by its RG-scale-$k$-dependent counterpart $G(k)$, and
	\item
	\textbf{(II) scale identification:} where the RG scale $k$ is identified with a physical scale such as, for instance, an appropriate power $n$ (i.e., fixed by dimensional analysis not involving $G_0$ itself) of a suitably chosen curvature invariant $\mathcal{K}$, i.e., $k\sim \mathcal{K}^n$.
\end{itemize}
Combining the two steps, RG-improvement defines a map $G_0\rightarrow G(\mathcal{K})$ from any classical spacetime to a distinct RG-improved spacetime.
At present, it is not conclusively known whether RG-improvement generates or even approximates solutions of some quantum or modified gravitational dynamics, cf.~\cite[VII.~D]{Bonanno:2020bil} as well as \cite{
tHooft:1984kcu,
AyonBeato:1998ub,
Nicolini:2008aj,
Gambini:2013ooa,
Haggard:2014rza,
Ashtekar:2018lag,
Nicolini:2019irw,
Bosma:2019aiu,
Borissova:2020knn,
Baake:2021jzv,
Chojnacki:2021ves
} for related works, as well as~\cite{
Poisson:1989zz,
Ori:1991zz,
Frolov:2017rjz,
Carballo-Rubio:2018pmi,
Carballo-Rubio:2019nel,
Bonanno:2020fgp,
Rubio:2021obb
} for aspects of stability and~\cite{
Borde:1996df,
Kumar:2019ohr,
Held:2019xde,
Contreras:2019cmf,
Liu:2020ola,
Zhou:2020eth,
Eichhorn:2021etc
} for phenomenological aspects of regular black holes, respectively. The present work will not add to these three crucial questions.
Instead, it focuses on the basic question of coordinate (in)dependence, i.e.,
\begin{quote}
	\emph{Does an RG-improvement procedure $G_0\rightarrow G(\mathcal{K})$, specified by (I) and (II), map a given classical spacetime to a unique RG-improved spacetime, irrespective of the coordinates in which the classical spacetime has been formulated?}
\end{quote}
Clarifying this question requires to determine whether two metrics are related by \emph{any} coordinate transformation. At the level of the metric (which is a tensor and its transformation thus includes derivatives of the coordinate transformation), this question amounts to solving differential equations. To circumvent this difficulty, the question can be approached by means of curvature invariants, cf.~\cite{Stephani:2003tm} for a collection of results in the context of GR. While the functional form of curvature invariants still transforms with a change of coordinates, their spacetime values, as well as all relations among them, are preserved. Hence, polynomial relations between invariants -- so-called syzygies -- can discriminate different spacetimes: Finding a particular combination of invariants that vanishes identically when evaluated on one spacetime while it does not vanish when evaluated on another spacetime, proves that the two spacetimes are inequivalent\footnote{
	The equivalence of two spacetimes is harder to establish and, in practice, requires (a) a complete set of scalar invariants that (b) is guaranteed to uniquely characterize the spacetimes at hand. Regarding (a), the literature distinguishes between functional and algebraic completeness. While the number of functionally independent scalar invariants is limited to at most the number of spacetime dimensions, the number of algebraically independent scalar invariants is known to be infinite~\cite{Stephani:2003tm}. Further distinguishing between Riemann invariants $\mathcal{K} = \lbrace R,R_{ab},R_{abcd}R^{abcd},\dots\rbrace$, i.e., the set of invariants formed purely by the Riemann tensor, and scalar polynomial invariants (SPI's) $\mathcal{I} = \lbrace R_{abcd}R^{abcd},\dots,R_{abcd;e}R^{abcd;e},\dots\rbrace$, i.e., the set of invariants formed by the Riemann tensor and its covariant derivatives. Clearly, $\mathcal{I}$ contains $\mathcal{K}$ and while for $\mathcal{K}$, a complete set is known~\cite{1991JMP....32.3135C, 1997GReGr..29..539Z, 2002nmgm.meet..831C}, this has not yet been achieved for $\mathcal{I}$. Regarding (b), it is known that not all spacetimes can be uniquely characterized in terms of $\mathcal{I}$~\cite{1961ZPhy..163...77K}. Those that can be, are referred to as $\mathcal{I}$-non-degenerate. In 4D, it is known that spacetimes are either $\mathcal{I}$-non-degenerate or of so-called Kundt-type~\cite{Coley:2009eb}. For $\mathcal{I}$-non-degenerate spacetimes, the Cartan-Karlhede algorithm~\cite{Cartan:1928,Karlhede:1979,Karlhede:1979ri,MacCallum:1993}, cf.~\cite[Ch.~9.2]{Stephani:2003tm} for a review, provides a way to unambiguously establish whether two metrics describe the same spacetime.
}.
Therefore, curvature invariants provide a means to settle the key question regarding coordinate transformations and RG improvement. 
\\

The rest of this work is structured as follows: Sec.~\ref{sec:synopsis} reviews the implementation of the physical input (I) and (II) for stationary black-hole spacetimes and summarizes the upshot of this work, which is simple and twofold:
On the one hand, cf.~Sec.~\ref{sec:metric-RG}, RG improvement at the level of the metric -- henceforth \emph{metric RG-improvement} -- is coordinate-dependent and thus not unique.
On the other hand, cf.~Sec.~\ref{sec:invariant-RG}, performing the RG-improvement at the level of curvature invariants -- henceforth \emph{invariant RG-improvement} -- can remedy the ambiguity and defines a coordinate-independent RG-improved spacetime. We conclude in Sec.~\ref{sec:discussion}. The review of an algebraically complete set of curvature invariants~\cite{1991JMP....32.3135C, 1997GReGr..29..539Z, 2002nmgm.meet..831C} and its application to the various black-hole spacetimes discussed in the main text is delegated to an appendix.

\section{Review \& synopsis}
\label{sec:synopsis}

Before giving a synopsis of the results, we briefly review the two physical inputs to RG-improvement, focusing on stationary GR black-hole spacetimes. The latter can be characterized through a complete set of curvature invariants, cf.~App.~\ref{app:ZM-basis} for a review of such an algebraically complete set. A characterization based on invariants is advantageous, compared to a metric description, not just because of coordinate-independence, but also because it is particularly simple. For instance, Kerr spacetime with mass $M$ and spin parameter $a$ is fully characterized by a single complex invariant (and the statement that other invariants vanish), i.e.,
\begin{align}
	\label{eq:II-Kerr}
	\mathbb{I} \equiv 
	C_{\mu\nu\rho\sigma}C^{\mu\nu\rho\sigma} + i\,C_{\mu\nu\rho\sigma}\overline{C}^{\mu\nu\rho\sigma}=
	\frac{48\,G_0^2M^2}{(r-i\,a \cos\theta)^6}\;,
\end{align}
where the explicit expression is given in ingoing Kerr coordinates\footnote{The expression also holds for Boyer-Lindquist (and any other set of) coordinates which are connected to ingoing Kerr coordinates by a transformation of the passive coordinates, i.e., those that do not appear in any invariants.}. In the above, $C$ denotes the Weyl tensor and $\overline{C}$ its dual.

\paragraph*{(I) Scale dependence.}
The first physical input, i.e., the dependence of, for instance, the Newton coupling $G$ on the RG scale $k$, arises naturally in the context of an asymptotically safe fixed point~\cite{Weinberg:1980gg,Reuter:1996cp}, cf.~also~\cite{Bonanno:2020bil} for a recent critical review. Herein, classical and quantum scale symmetry unambiguously set the two respective limit behaviours $G(k)\stackrel{k\rightarrow 0}{\longrightarrow} G_0$ and $G(k)\stackrel{k\rightarrow \infty}{\longrightarrow} \sim k^{-2}$.
These limits can be interpolated by
\begin{align}
	\label{eq:scale-dependence}
	G(k) = \frac{G_0}{1 + \ell_\text{NP}^2k^2}\;,
\end{align}
where the new-physics scale (or equivalently the Planck length) $\ell_\text{NP}^2 = G_0/g_\ast$ sets the transition scale. This scale is set by (i) the measured low-energy value of the \emph{dimensionful} Newton coupling $G_0$ and (ii) the fixed-point value of the \emph{dimensionless} Newton coupling $g_\ast$. (We work in natural units in which $c\equiv\hbar\equiv 1$.) The above RG-scale dependence also arises from integration of the $\beta$-function of the Newton coupling, when neglecting the running of all other couplings~\cite{Bonanno:1998ye}. 
In fact, other couplings (such as the cosmological constant or higher-curvature couplings) are scale-dependent as well and may thus also be RG-improved. Here, we will restrict to RG-improving the Newton coupling but the main conclusions regarding coordinate (in)dependence apply more generally.
Beyond the above limits, $G(k)$ is known in functional-RG~\cite{Wetterich:1992yh} truncations of the quantum effective action, cf.~\cite{Percacci:2017fkn,Eichhorn:2018yfc,Reuter:2019byg,Reichert:2020mja} for textbooks and recent reviews in the context of asymptotic safety.

\paragraph*{(II) Scale identification.}
The second physical input, i.e., the scale identification, can be tied to the physical principle that quantum fluctuations (or new physics more generally) set in beyond some critical \emph{local} curvature scale, cf.~\cite{Eichhorn:2021etc,Eichhorn:2021iwq}. Following~\cite{Eichhorn:2021iwq}, the latter can be meaningfully defined as the maximum of the absolute value of \emph{all} (algebraically) independent curvature invariants $\mathcal{K}_n$ (see App.~\ref{app:ZM-basis} for such a complete set), i.e.,
\begin{align}
	k^D \sim \max_n\left\lbrace (|\mathcal{K}_n|^{m_n}\right\rbrace\;,
\end{align}
where the exponent $m_n$ is set such that the mass dimension of $(\mathcal{K}_n)^{m_n}$ equals the spacetime dimension $D$. 
For Kerr spacetime in $D=4$, being characterized by a single complex invariant, cf.~Eq.~\eqref{eq:II-Kerr}, this motivates, cf.~\cite{Eichhorn:2021iwq},
\begin{align}
\label{eq:scale-identification}
k^4 = |\mathbb{I}| = \frac{48\,G_0^2M^2}{(r^2 + a^2\cos^2\theta)^3}\;,
\end{align}
Whenever an explicit RG-improvement is required, we follow this physical input.
In spherical symmetry, this scale-identification reduces to the Kretschmann scalar and therefore reproduces previous RG-improvements of Schwarzschild spacetime~\cite{Bonanno:1998ye} in the limit of vanishing spin parameter, i.e., $a\rightarrow 0$.

\paragraph*{Result: Metric RG-improvement is coordinate-dependent.} In the context of black holes, RG-improvement has mostly been performed at the level of the metric -- sometimes referred to as RG-improvement at the level of solutions. However, see e.g.~\cite{Bonanno:2001hi,Bonanno:2002zb,Babic:2004ev} and \cite{Reuter:2003ca,Reuter:2004nv} for studies in cosmology with an RG-improvement at the level of the equations of motion and at the level of the action, respectively. Here, we focus on RG improvement at the level of solutions. Utilizing polynomial relations among curvature invariants, we unambiguously prove, cf.~Sec.~\ref{sec:metric-RG}, that 
\begin{itemize}
	\item
	the resulting metric-RG-improved spacetime depends on the coordinates in which the classical spacetime is expressed, i.e., metric-RG improvement and coordinate transformations do not commute, cf.~Fig.~\ref{fig:non-comm-metric-RG};
	\item
	different horizon-penetrating coordinates result in physically distinct metric-RG-improved spacetimes which appear equally viable, i.e., do not exhibit any curvature singularities;
	\item
	coordinate singularities can induce novel curvature singularities in metric-RG-improved spacetimes.
\end{itemize}
Explicit examples in spherical as well as axial symmetry are given in Sec.~\ref{sec:metric-RG}. Spinning black holes are of particular astrophysical interest: using (I) the scale-dependence in Eq.~\eqref{eq:scale-dependence} and (II) the scale-identification in Eq.~\eqref{eq:scale-identification} to replace $G_0\rightarrow G(k(\mathcal{K}))$ in horizon-penetrating ingoing Kerr coordinates, cf.~Eq.~\eqref{eq:lineElem-Kerr}, results in a member of the spacetimes phenomenologically characterized in~\cite{Eichhorn:2021etc,Eichhorn:2021iwq}. In contrast, describing the classical spacetime in Boyer-Lindquist coordinates, cf.~\cite{Held:2019xde}, results in a metric-RG-improved spacetime with curvature singularities at the black-hole horizon.

Similar conclusions about coordinate-dependence are expected to also hold for RG improvement at the level of the equations of motion, cf.~\cite{Bonanno:2001hi,Bonanno:2002zb,Babic:2004ev}, where the RG improvement is also used to replace non-scalar quantities. 
In contrast, RG improvement at the level of the action, cf.~\cite{Reuter:2003ca,Reuter:2004nv}, involves scalar quantities only and, hence, is expected to be coordinate-independent, see below.

\paragraph*{Result: Invariant RG-improvement is coordinate independent.} A novel coordinate independent RG-improvement procedure at the level of solutions can be achieved if RG-improvement is performed, not for the metric, but rather for the curvature invariants that characterize the classical solution, cf.~Sec.~\ref{sec:invariant-RG} and Fig.~\ref{fig:comm-inv-RG}.
In summary, invariant RG-improvement amounts to:
\begin{itemize}
	\item
	specifying (I) scale dependence and (II) scale identification to give $G(k(\mathcal{K}_i))$, as reviewed above;
	\item
	identifying a suitable (functionally independent) set of curvature invariants $\mathcal{K}_i$;
	\item
	replacing $G_0\rightarrow G(k(\mathcal{K}_i))$ in the $\mathcal{K}_i$ themselves to obtain the RG-improved invariants.
\end{itemize}
Optionally, one may use a general metric ansatz and integrate the resulting set of differential equations to reconstruct the respective invariant-RG-improved metric. In practice, this last step will typically be non-trivial as it involves solving (coupled) differential equations.

\section{Metric Renormalization-Group improvement}
\label{sec:metric-RG}

This section delineates the dependence of metric RG-improvement on the coordinates in which the classical metric is formulated. In hindsight, such dependence may not be surprising since metric components themselves are coordinate-dependent quantities. Given a classical spacetime in two sets of coordinates, Sec.~\ref{sec:non-commute} gives a generic argument that the two RG-improved spacetimes are related neither by the classical nor by a `na\"ively RG-improved' coordinate transformation.

Since this does not conclusively disprove that the two RG-improved spacetimes are related by some other coordinate transformation, explicit examples for Schwarzschild and Kerr black holes are given in Sec.~\ref{sec:sph-symm} and \ref{sec:axial-symm}, respectively. For these physically relevant cases, it is unambiguously proven that distinct RG-improved spacetimes, not related by \emph{any} coordinate transformation, can be obtained. These examples also showcase that the choice of coordinates is decisive to even the most fundamental properties of the metric-RG-improved spacetime such as whether or not curvature singularities exist.

\subsection{Metric-RG improvement and coordinate transformations}
\label{sec:non-commute}

\begin{figure}
	\begin{diagram}[heads=LaTeX]
		g(G_0,X) &
		\rTo^{\text{coordinate trafo}}_{X\mapsto \underline{X} = F(X)}   &
		g(G_0,\underline{X})
	\\
		\dTo_{\text{RG}}  &
		&
		\dTo_{\text{RG}}
	\\
		\widetilde{g}(G(\mathcal{K}(X)),X) &
		\neq &
		\underline{\widetilde{g}}(G(\mathcal{K}(\underline{X})),\underline{X})
	\end{diagram}
	\caption{\label{fig:non-comm-metric-RG}Noncommutative diagram for metric RG-improvement and coordinate transformations.}
\end{figure}

Consider a classical spacetime, characterized by a metric $g$ or, equivalently, by a complete set of curvature invariants $\mathcal{K}$. The metric or the invariants may be described in two sets of coordinates $X_a$ and $\underline{X}_{\underline{a}}$ which are related by a coordinate transformation $F$, i.e.,
\begin{align}
	F:X^a\mapsto\underline{X}^{\underline{a}} = F^{\underline{a}}(X)\;.
\end{align}
In particular, while the curvature invariants transform as scalars, i.e., $\mathcal{K}(X) = \mathcal{K}(\underline{X}(X))$, the metric transforms as a tensor. Put differently, the invariant line element $ds^2$ may be expressed in both coordinates as
\begin{align}
	\text{ds}^2 = 
	g_{\underline{a}\underline{b}}(\underline{X})\;
	\text{d}\underline{X}^{\underline{a}}\,\text{d}\underline{X}^{\underline{b}} = 
	g_{\underline{a}\underline{b}}(\underline{X})\;
	\frac{\partial \underline{X}^{\underline{a}}}{\partial X^a}
	\frac{\partial \underline{X}^{\underline{b}}}{\partial X^b}
	\;\text{d}X^a\,\text{d}X^b =
	g_{\underline{a}\underline{b}}(F(X))\;
	f^{\underline{a}}_{\phantom{a}a}\,f^{\underline{b}}_{\phantom{b}b}\;
	\text{d}X^a\,\text{d}X^b\;.
\end{align}
Crucially, the transformation of the line element (and equivalently the metric) involves both the coordinate transformation $F$ as well as its derivatives $f^{\underline{a}}_{\phantom{a}a} = \partial F^{\underline{a}}(X)/\partial X^a$. (The separate notation helps with clarity in the following discussion.)
\\

As revewed in the Sec.~\ref{sec:synopsis}, metric RG-improvement amounts to improving $G_0$ in the metric components via
\begin{align}
\label{eq:RG-improvement}
	G_0\longrightarrow G(X)
	\quad\quad\quad\text{or}\quad\quad\quad
	G_0\longrightarrow G(\underline{X})\;,
\end{align}
in either coordinates.
If and only if $G(\mathcal{K}(X))$ is a function only of curvature scalars of the classical spacetime, then $G(\mathcal{K}(\underline{X}))$ remains a curvature scalar with respect to the old spacetime. This will be of importance concerning invariant RG-improvement in Sec.~\ref{sec:invariant-RG}. However, it is neither necessary nor sufficient for the considerations of the present section, i.e., regarding metric RG-improvement.

Whenever $G_0$ appears in the classical coordinate transformation $F(X,G_0)$ as well as its derivatives $f^{\underline{a}}_{\phantom{a}a}(X,G_0)$, noncommutativity of metric RG-improvement and coordinate transformations, cf.~Fig.~\ref{fig:non-comm-metric-RG}, is apparent. On the one hand, the classical transformation can no longer relate the two RG-improved metrics because one has replaced $G_0\rightarrow G$ in the metric components. On the other hand, replacing $G_0$ also in $F(X,G)$ and $f^{\underline{a}}_{\phantom{a}a}(X,G)$ upsets the relation of the latter two: $f^{\underline{a}}_{\phantom{a}a}(X,G)$ is no longer the Hessian of $F(X,G)$, i.e.,
\begin{align}
	f^{\underline{a}}_{\phantom{a}a}(X,G(X)) \neq \frac{\partial F^{\underline{a}}(X,G(X))}{\partial X^a}\;.
\end{align}
Hence, neither the classical coordinate transformation nor the `na\"ive' RG-improved coordinate transformation attempted above relate the two metric-RG-improved spacetimes.

Although this makes the problem manifest, it does not conclusively disprove the existence of \emph{some} other coordinate transformation. Therefore, important explicit examples will be discussed below. For some pairs of coordinate choices, it can be unambiguously established that the two metric-RG-improved spacetimes are not be related by \emph{any} coordinate transformation.

\subsection{Black holes in spherically symmetry}
\label{sec:sph-symm}
As a first explicit example, consider metric RG-improvement of a Schwarzschild black-hole with mass $M$. The classical line element in ingoing Eddington-Finkelstein (EF) coordinates $\lbrace u,r,\theta,\phi\rbrace$ is given by
\begin{align}
\label{eq:lineElem_EF}
	\text{ds}_\text{EF} =& 
	-\left(1-\frac{2MG_0}{r}\right)\text{d$u$}^2
	+ 2\,\text{d$u$}\,\text{d$r$}
	+r^2(\text{d$\theta$}^2+\sin^2\theta\,\text{d$\phi$}^2)\;.
\end{align}
These coordinates are horizon-penetrating and free of coordinate singularities. Alternatively, the classical metric may be transformed into Schwarzschild coordinates $\lbrace t,r,\theta,\phi \rbrace$, for which the corresponding line element and transformation read
\begin{align}
\label{eq:lineElem_Schwarzschild}
	\text{ds}_\text{Schw} &= 
	-\left(1-\frac{2MG_0}{r}\right)\text{d$t$}^2
	+\left(1-\frac{2MG_0}{r}\right)^{-1}\text{d$r$}^2
	+r^2(\text{d$\theta$}^2+\sin^2\theta\,\text{d$\phi$}^2)\;,
	\\[0.5em]
\label{eq:trafo-EF-Schw}
	t &= u - r - 2MG_0\log(r-2MG_0)
	\quad\Longrightarrow\quad
	\text{d$u$} = \text{d$t$} + \left(1-\frac{2MG_0}{r}\right)^{-1} \text{d$r$}\;.
\end{align}
In contrast to EF coordinates, Schwarzschild coordinates are singular at the horizon, i.e., at $r=2MG_0$. For exemplary purposes, we define yet another set of coordinates $\lbrace u,x,\theta,\phi\rbrace$. These are less conventional but can be obtained from EF coordinates, i.e.,
\begin{align}
\label{eq:lineElem_EF-mod}
	\text{ds}_\text{EF-mod} &= 
	-\left(1-\frac{2(MG_0)^n}{x^n}\right)\text{d$u$}^2
	+ \frac{2n\,x^{n-1}}{(MG_0)^{n-1}}\,\text{d$u$}\text{d$x$}
	+\frac{x^{2n}}{(MG_0)^{2(n-1)}}(\text{d$\theta$}^2+\sin^2\theta\,\text{d$\phi$}^2)\;,
	\\
\label{eq:trafo-EF-ntic}
	x &= \frac{r^n}{(MG_0)^{n-1}}
	\quad\Longrightarrow\quad
	\text{d$x$} = \frac{n\,r^{n-1}}{(MG_0)^{n-1}}\text{d$r$}\;,
\end{align}
with $n\in\mathbb{N}^+$.
The latter serve as a second example for horizon-penetrating coordinates.

Naturally, all relations between curvature invariants $\mathcal{K}$ of the classical Schwarzschild spacetime are preserved, independent of the choice of coordinates. Due to spherical symmetry, the invariants are a function of the respective radial coordinate only, i.e., $\mathcal{K}(r)$ (for Schwarzschild and EF coordinates) or $\mathcal{K}(x)$ (for modified-EF coordinates). The classical spacetime is metric RG-improved by replacing 
\begin{align}
	G_0\rightarrow G(\mathcal{K})\equiv G(r) \equiv G(x)\;,
\end{align}
for each choice of coordinates, respectively. The explicit function $G$ need not be specified in the following but one may use the physical input specified in Sec.~\ref{sec:synopsis}. When evaluated on the three resulting spacetimes, the following Ricci invariants can explicitly be calculated as functions of $G(r)$ or $G(x)$, respectively:
\begin{align}
	\mathcal{K}_5 = g_{ab}R^{ab}\;,
	\quad
	\mathcal{K}_6 = R_{ab}R^{ab}\;,
	\quad
	\mathcal{K}_7 = R_{a}^{b}\;R_{b}^{c}\;R_{c}^{a}\;,
	\quad
	\mathcal{K}_8 = R_{a}^{b}\;R_{b}^{c}\;R_{c}^{d}\;R_{d}^{a}\;.
\end{align}
All four invariants are zero on the classical Schwarzschild spacetime (due to Ricci-flatness) but no longer vanish on any of the metric RG-improved spacetimes. The complete set of ZM-invariants and their explicit expressions in each of the three cases may be found in App.~\ref{app:ZM_metric-RG_spherical} and in the ancillary files~\cite{anc}. Given these invariants, one can construct two combined polynomial invariants
\begin{align}
\label{eq:syzygy1}
	\mathfrak{K}_1 &= \frac{1}{8}\left(\mathcal{K}_5^2 - 2\,\mathcal{K}_6\right)^2 - \left(\mathcal{K}_6^2 - 2\,\mathcal{K}_8\right)\;,
	\\
\label{eq:syzygy2}
	\mathfrak{K}_2 &= \frac{1}{8}\mathcal{K}_5\left(\mathcal{K}_5^2 - 6\,\mathcal{K}_6\right)^2 +\mathcal{K}_7\;.
\end{align}
After metric RG-improvement, $\mathfrak{K}_1(r)\equiv 0$ and $\mathfrak{K}_2(r)\equiv 0$ vanish in EF as well as in Schwarzschild coordinates, but $\mathfrak{K}_1(x)\neq 0$ and $\mathfrak{K}_2(x)\neq 0$ in modified-EF coordinates. This establishes that the latter metric RG-improved spacetime is inequivalent to the former two. The other two metric RG-improved spacetimes (originating from EF and Schwarzschild coordinates) are indeed equivalent, as addressed below in terms of an explicit transformation and in App.~\ref{app:ZM_metric-RG_EF} in terms of a complete set of invariants.
\\

A closer look at the two transformations from EF to Schwarzschild in Eq.~\eqref{eq:trafo-EF-Schw} and from EF to modified-EF in Eq.~\eqref{eq:trafo-EF-ntic} reveals: the latter requires (ultimately conflicting) relations at the coordinate and the differential level; the former only requires a specific form of differential relation. 

This occurs because, in case of the transformation from EF to Schwarzschild, the transformed coordinate ($u\leftrightarrow t$) is absent in the metric components. When replacing $G_0\rightarrow G(r)$ in the differential relation in Eq.~\eqref{eq:trafo-EF-Schw}, the latter still defines a proper coordinate transformation. Indeed, the two metric RG-improved spacetimes -- rooted in EF and Schwarzschild coordinates -- are related by
\begin{align}
	\text{d$u$} = \text{d$t$} + \left(1-\frac{2MG(r)}{r}\right)^{-1} \text{d$r$}\;.
\end{align}
If $G(r)$ is non-trivial, integrating this differential relation will no longer result in the coordinate relation in Eq.~\eqref{eq:trafo-EF-Schw} but rather in some more complicated transformation. However, the integrated form is insignificant for the question of whether the transformation relates the two RG-improved metrics because the transformed coordinate does not appear in the metric components.
\\

Coming back to the transformation between EF and modified-EF coordinates in Eq.~\eqref{eq:trafo-EF-ntic}, the differential and the coordinate relation are inconsistent for non-trivial $G(r)$, and, in contrast to the previous example (between Schwarzschild and EF coordinates), both relations need to be maintained to successfully transform between the two line elements. Hence, Eq.~\eqref{eq:trafo-EF-ntic} does no longer correspond to a coordinate transformation after metric RG-improvement. Moreover, we have established the inequivalence of the curvature relations $\mathfrak{K}_1$ and $\mathfrak{K}_2$ on the respective spacetimes.

While physically distinct, both spacetimes appear equally viable. In particular, for sufficiently weakened $G(r\ll M)\sim r^{m<-3}$ (and an equivalent scaling for $G(x\ll M)$) both spacetimes are regular at the center, i.e., at $r=0$ ($x=0$). There is no physical argument to prefer either of the two.

\subsection{Black holes in axial symmetry}
\label{sec:axial-symm}

Reducing the symmetry from static and spherical to stationary and axial, the Kerr metric exemplifies that in less symmetric settings even the most conventional coordinate choices can lead to very different physical conclusions when metric-RG-improving the spacetime. 
Ingoing Kerr coordinates $\lbrace u,r,\theta,\varphi\rbrace$ provide the axisymmetric generalization of ingoing EF coordinates, i.e.,
\begin{align}
\label{eq:lineElem-Kerr}
	\text{d$s$}_\text{Kerr} =&
	-\frac{\left(\Delta -a^2 \sin^2\theta\right)}{\Sigma }\text{du}^2
	+2\,\text{du}\text{dr}
	-\frac{2 a \sin^2\theta \left(a^2-\Delta +r^2\right)}{\Sigma }\text{du}\text{d$\varphi $}
	-2 a \sin^2\theta \text{dr} \text{d$\varphi $}
	\notag\\&
	+\Sigma\,\text{d$\theta $}^2
	+\frac{\sin^2\theta \left(\left(a^2+r^2\right)^2-a^2 \Delta  \sin^2\theta\right)}{\Sigma } \text{d$\varphi $}^2
	\;,
\end{align}
where $\Delta = r^2 - 2M G_0 r + a^2$ and $\Sigma = r^2 + a^2\cos^2\theta$. Moreover, $M$ and $a$ denote the mass and the spin parameter of the black hole, respectively. In particular, ingoing Kerr-coordinates are horizon-penetrating and exhibit no coordinate singularities. 

Alternatively, the Kerr metric may be expressed in Boyer-Lindquist (BL) coordinates $\lbrace t,r,\theta,\phi\rbrace$, i.e.,
\begin{align}
\label{eq:lineElem-BL}
	\text{d$s$}_\text{BL}=& 
	-\frac{\Delta - a^2\,\sin(\theta)^2}{\Sigma}\,dt^2
	+\frac{\Sigma}{\Delta}\,dr^2
	+\rho^2\,d\theta^2
	+\frac{(a^2+r^2)^2 - a^2\,\Delta\,\sin(\theta)^2}{\Sigma}\sin(\theta)^2\,d\phi^2
	\nonumber\\&
	-\frac{2(a^2+r^2-\Delta)}{\Sigma}a\,\sin(\theta)^2\,dt\,d\phi\;.
\end{align}
These are related to ingoing Kerr coordinates via the coordinate transformation
\begin{align}
\label{eq:trafo-Kerr-BL}
	du &= dt + \frac{r^2 + a^2}{\Delta}dr\;,
	\quad\quad\quad
	d\varphi = d\phi - \frac{a}{\Delta}dr\;.
\end{align}
BL coordinates manifestly exhibit asymptotic flatness. Further, their existence is related to an `accidental' Killing tensor (in addition to the two Killing vectors associated with stationarity and axisymmetry), the existence of a third constant of motion -- the Carter constant~\cite{Carter:1968rr, Carter:1971zc} -- and the resulting separability of the Hamilton-Jacobi and scalar-wave equations, see, e.g.,~\cite{Teukolsky:2014vca}.
However, BL coordinates feature a coordinate singularity at the event horizon, i.e., at $\Delta = 0$.
\\

As for the spherically-symmetric case, RG-improvement amounts to the replacement
\begin{align}
	G_0 \rightarrow G(\mathcal{K}) \equiv G(r,\theta)\;,
\end{align}
where $\mathcal{K}$ is some combination of curvature invariants, all of which only depend on $r$ and $\theta$ due to axial symmetry and stationarity. Crucially, $G_0$ appears in the coordinate transformation between ingoing Kerr and BL coordinates in Eq.~\eqref{eq:trafo-Kerr-BL}. Therefore, when metric-RG improving $G_0 \rightarrow G(r,\theta)$, Eqs.~\eqref{eq:trafo-Kerr-BL} cease to describe exact differentials due to the newly introduced $\theta$-dependence. This makes apparent, that the two metric-RG-improved spacetimes are no longer related by a coordinate transformation -- at least not by the naively attempted one described above.
\\

The above expectation can be confirmed by calculating the Ricci scalar $\mathcal{K}_5 = g_{ab}R^{ab}$
-- note that the improved spacetimes are no longer Ricci flat -- evaluated on the respective metric-RG-improved spacetimes, i.e.,
\begin{align}
	\label{eq:K5-Kerr}
	\mathcal{K}_{5,\,\text{Kerr}} &\!=\! \frac{4 M \left(2  G^{(1,0)}(r,\theta )+r  G^{(2,0)}(r,\theta )\right)}{a^2 \cos (2 \theta )+a^2+2 r^2}\;,
	\\[0.5em]
    \label{eq:K5-BL}
	\mathcal{K}_{5,\,\text{BL}} &\!=\! \frac{2M}{r^2 + a^2\cos(\theta)^2}\left[
		2G^{(1,0)}(r,\theta) + r\!\left(G^{(2,0)}(r,\theta) + \frac{Mr(\cos(\theta)^2-1)G^{(0,1)}(r,\theta)^2}{\left(r^2 - 2M G(r,\theta) r + a^2\right)^2}\right)
	\!\right]\,.
\end{align}
where the superscripts $^{(n_r,n_\theta)}$ denote $n_r$ and $n_\theta$ partial derivatives in $r$- and $\theta$-direction, respectively.

On the one hand, since classical ingoing Kerr and BL coordinates are related by transformations of only the passive coordinates, i.e., the active coordinates $r$ and $\theta$ remain untransformed, one may be inclined to already conclude that the respective Ricci scalars in Eqs.~\eqref{eq:K5-Kerr}-\eqref{eq:K5-BL} above do \emph{not} agree whenever $G(r,\theta)$ is non-trivial.

On the other hand, one may object that the coordinates called $r$ and $\theta$, respectively, in each of the expressions in Eqs.~\eqref{eq:K5-Kerr}-\eqref{eq:K5-BL}, do not necessarily have to agree: in principle, the RG-improved spacetimes may become related by a new coordinate transformation which includes a non-trivial transformation of these active coordinates. 
This loophole can be closed by checking the syzygies in Eqs.~\eqref{eq:syzygy1}-\eqref{eq:syzygy2}. These vanish identically for the metric-RG-improved spacetime rooted in ingoing Kerr coordinates, cf.~App.~\ref{app:ZM_metric-RG_ingoingKerr}, but remain non-vanishing for metric-RG-improved spacetime rooted in BL coordinates, cf.~App.~\ref{app:ZM_metric-RG_BL}. This conclusively establishes that the two spacetimes are physically distinct.
\\

Moreover, $\mathcal{K}_{5,\,\text{BL}}$ exhibits a curvature singularity at $\left(r^2 - 2M G(r,\theta) r + a^2\right)^2 = 0$, even if $G(r,\theta)$ as well as its derivatives are analytic. This exemplifies that metric RG-improvement can turn coordinate singularities into curvature singularities. In contrast, when evaluated on the ingoing-Kerr metric-RG-improved spacetime, $\mathcal{K}_{5,\,\text{Kerr}}$, as well as all other curvature invariants, cf.~App.~\ref{app:ZM_metric-RG_ingoingKerr}, are regular at the horizon. This provides an example that, regarding metric RG-improvement, even basic questions like the regularity of the resulting spacetime are coordinate-dependent statements.

\section{Invariant Renormalization-Group improvement}
\label{sec:invariant-RG}

The coordinate-dependence of RG-improvement at the metric level arises because the metric itself transforms as a tensor and not as a scalar. Its components have no direct physical meaning. Naturally, RG-improving the latter is \emph{not} a coordinate-invariant procedure. 

Instead, RG-improvement may be performed at the level of curvature invariants which do transform as scalars. Let $\mathcal{K}_i$ be such a set of invariants and $X$ as well as $\underline{X} = F(X)$ two sets of classical coordinates connected by a coordinate transformation $F: X\mapsto\underline{X}$. The $\mathcal{K}_i$ may represent an algebraically (or functionally) complete set of invariants~\cite{1991JMP....32.3135C, 1997GReGr..29..539Z, 2002nmgm.meet..831C}, cf.~App.~\ref{app:ZM-basis} or, in fact, any other set of curvature invariants. Invariant RG-improvement can then be defined as the replacement
\begin{align}
	\mathcal{K}_i(G_0,X)&\rightarrow \widetilde{\mathcal{K}}_i(G(\mathcal{K}_j(X)),X)\;,
	\\
	\mathcal{K}_i(G_0,\underline{X})&\rightarrow \widetilde{\mathcal{K}}_i(G(\mathcal{K}_j(\underline{X})),\underline{X})\;,
\end{align}
where crucially, $G(\mathcal{K}_j)$ is a function that only depends on the curvature invariants of the classical spacetime, cf.~Sec.~\ref{sec:synopsis}.
Since the classical curvature invariants $\mathcal{K}_i(X) = \mathcal{K}_i(\underline{X}(X))$ transform as scalars under the classical coordinate transformation, so do the RG-improved ones (denoted by an added $\widetilde{\phantom{a}}$), i.e.,
\begin{align}
	\widetilde{\mathcal{K}}_i(G(\mathcal{K}_j(X)),X) = \widetilde{\mathcal{K}}_i(G(\mathcal{K}_j(\underline{X}(X))),\underline{X}(X))\;.
\end{align}
In particular, all (algebraic and functional) relations between the improved invariants $\widetilde{\mathcal{K}}_i$ will thus be preserved. Hence, invariant RG-improvement and coordinate transformations commute, i.e., the RG-improved invariants are related by the same coordinate transformation $F: X\mapsto\underline{X}$ as the classical invariants, cf.~Fig.~\ref{fig:comm-inv-RG}. Assuming $\mathcal{I}$-non-degeneracy, cf.~\cite{Coley:2009eb}, the two sets of invariants therefore describe the same invariant-RG-improved spacetime.
\\

\begin{figure}
	\begin{diagram}[heads=LaTeX]
		\mathcal{K}(G_0,X) &
		\rTo^{\text{coordinate trafo}}_{X\mapsto \underline{X} = F(X)}   &
		\mathcal{K}(G_0,\underline{X})
	\\
		\dTo_{\text{RG}}  &
		&
		\dTo_{\text{RG}}
	\\
		\widetilde{\mathcal{K}}(G(\mathcal{K}(X)),X) &
		\rTo^{\text{coordinate trafo}}_{X\mapsto \underline{X} = F(X)} &
		\widetilde{\mathcal{K}}(G(\mathcal{K}(\underline{X})),\underline{X})
	\end{diagram}
	\caption{\label{fig:comm-inv-RG}Commutative diagram for invariant RG-improvement and coordinate transformations.}
\end{figure}

We caution that it is not a priori clear under which conditions the resulting set of RG-improved invariants corresponds to a consistent spacetime. The proof in~\cite{Coley:2009eb} guarantees that any $\mathcal{I}$-non-degenerate spacetime is fully characterized by its invariants. However, this does not ensure that a generic deformation of such a set of characterizing invariants -- and RG-improvement in particular -- results in a new set of invariants that remain connected to a correspondingly deformed spacetime.

It seems reasonable to conjecture that (i) a functionally complete set of invariants and (ii) a smooth deformation parameter are necessary (and potentially also sufficient) conditions for the existence of a corresponding spacetime. Proof of this claim remains for future work. In the context of spherically-symmetric black holes, cf.~Sec.~\ref{sec:invRG-Schw} below, we work with an algebraically complete set of invariants and find that inconsistencies arise if one enforces particular invariants to remain vanishing.
\\

If the given set of RG-improved invariants does not correspond to any spacetime, invariant-RG-improvement may thus have to be supplemented by a prescription of which invariant relations should be enforced and which ones not. This would correspond to an additional physical choice, namely to select which particular part of the algebraically special character of the classical spacetime should be broken. That said, we leave a more complete analysis of this consistency question for future work.
\\

Provided the RG-improved invariants $\widetilde{\mathcal{K}}_i$ correspond to a consistent spacetime, reconstructing the respective metric remains a non-trivial task. One potential ansatz to do so is to calculate the invariants $\widehat{\mathcal{K}}_i$ of a generic spacetime metric $\widehat{g}$ that obeys the symmetries of the problem. Solving the resulting set of differential equations, i.e.,
\begin{align}
	\widetilde{\mathcal{K}}_i = \widehat{\mathcal{K}}_i\;,
\end{align}
amounts to reconstructing the invariant-RG-improved spacetime metric.
\\

In the remainder of this section, the procedure of invariant-RG-improvement is demonstrated for Schwarzschild spacetime. Furthermore, we outline its application to Kerr spacetime.

\subsection{Invariant-RG-improved Schwarzschild spacetime}
\label{sec:invRG-Schw}

Due to staticity and spherical symmetry, Schwarzschild spacetime is characterized by a single real-valued curvature invariant (and the demand that the other ones vanish\footnote{Integrating $\mathcal{K}_1$ to obtain $A(r) \equiv B(r)$ in the ansatz in Eq.~\eqref{eq:sphSymm-ansatz} results in $A(r) = 1-2MG_0/r + c_1 r + c_2 r^2$ with $c_1$ and $c_2$ some integration constants. Only by additionally integrating, for instance, $\mathcal{K}_5=0$, which results in $A(r) = 1 + c_3/r + c_4/r^2$, one finds $c_1=c_2=0$.}), i.e., in terms of ZM-invariants~\cite{1991JMP....32.3135C, 1997GReGr..29..539Z, 2002nmgm.meet..831C}, cf.~App.~\ref{app:ZM-basis} and~\ref{app:ZM-classical},
\begin{align}
	\left(\frac{\mathcal{K}_1}{48}\right)^3 = \left(\frac{\mathcal{K}_3}{96}\right)^2 = \left(\frac{G_0\,M}{r^3}\right)^6\;,
	\quad\quad\quad
	\mathcal{K}_{i\neq1,3} = 0\;.
\end{align}
Due to the high degree of symmetry, the most general RG-improvement can be expressed as $G_0\rightarrow G(\mathcal{K}_1(r))$ and the improved invariants are given by
\begin{align}
	\widetilde{\mathcal{K}}_1 = 48\left(\frac{G(\mathcal{K}_1(r))M}{r^3}\right)^2\;,
	\quad\quad\quad
	\widetilde{\mathcal{K}}_3 = 96\left(\frac{G(\mathcal{K}_1(r))M}{r^3}\right)^3\;,
	\quad\quad\quad
	\widetilde{\mathcal{K}}_{i\neq1,3} = 0\;.
\end{align}
We emphasize again that this is a coordinate-invariant procedure and that all relations among the invariants are preserved. 
\\

Birkhoff's theorem -- and the Ricci-flatness which is implied by $\widetilde{\mathcal{K}}_{i\neq1,3} = 0$ -- immediately lead to the conclusion that Schwarzschild spacetime is the only consistent solution. Vice versa, for any non-constant $G(\mathcal{K}_1(r))$, the attempt to enforce the full set of ZM invariants -- including the vanishing ones -- does not correspond to a consistent spacetime. Hence, we know in advance that we must drop part of the algebraically special character, i.e., drop (some of) the $\widetilde{\mathcal{K}}_{i\neq1,3} = 0$ relations. 

On the one hand, it is in agreement with general expectations that RG-improvement has to break Ricci-flatness. On the other hand, it exemplifies that while invariant RG-improvement is manifestly coordinate-independent, it might require a subsequent physical choice about how to reduce the algebraically special character. The main advantage, in comparison to coordinate-dependent metric RG-improvement, reviewed in Sec.~\ref{sec:metric-RG}, is that such an additional choice can be implemented based on physical principles rather than an unphysical, and hence uncontrollable, influence of the choice of coordinates.
For example, we choose to drop the requirement of Ricci-flatness ($\widetilde{\mathcal{K}}_{i\geqslant5} = 0$), in the following.
\\

\begin{figure}
	\centering
	\includegraphics[width=0.48\linewidth]{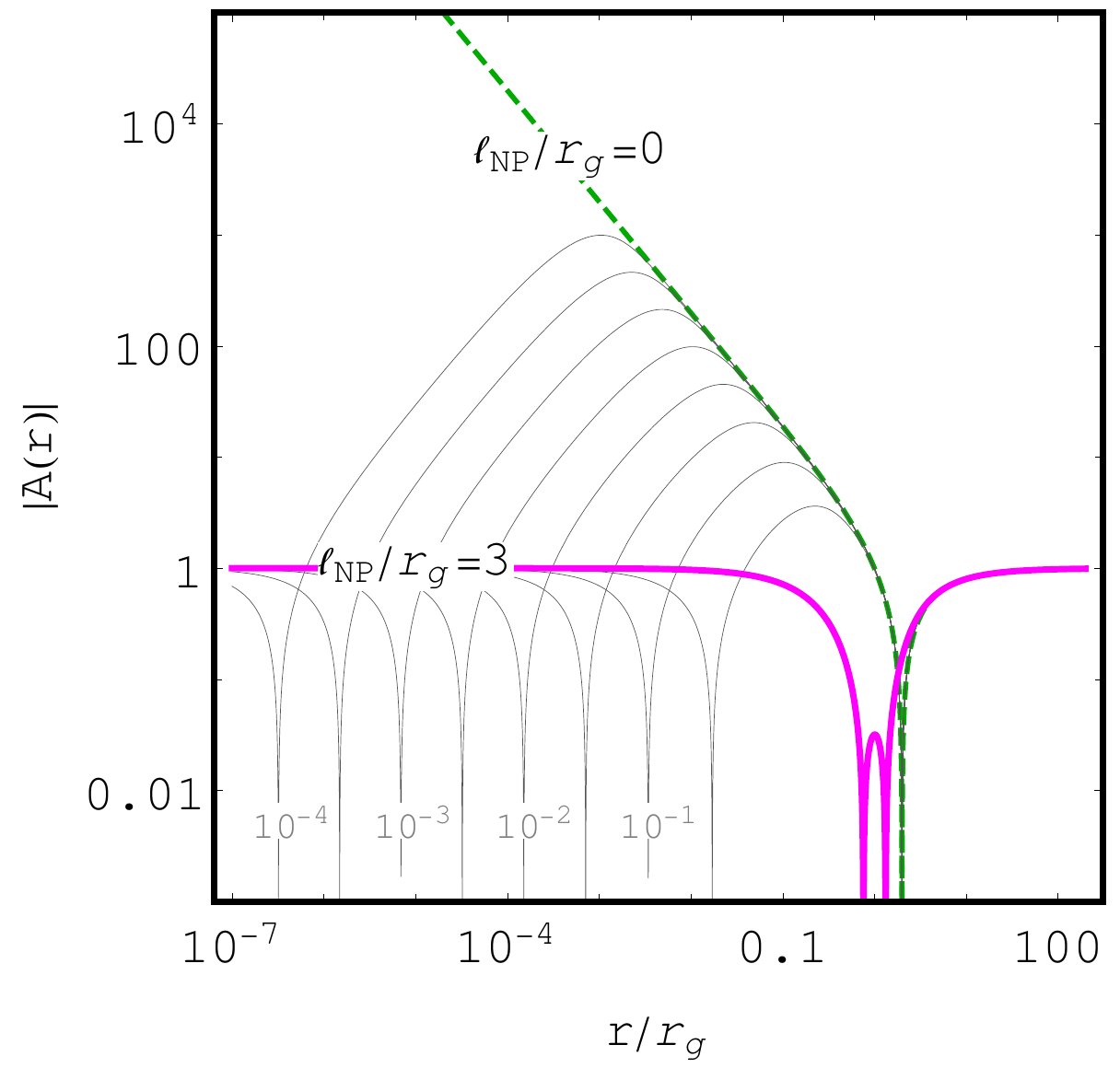}
	\hfill
	\includegraphics[width=0.48\linewidth]{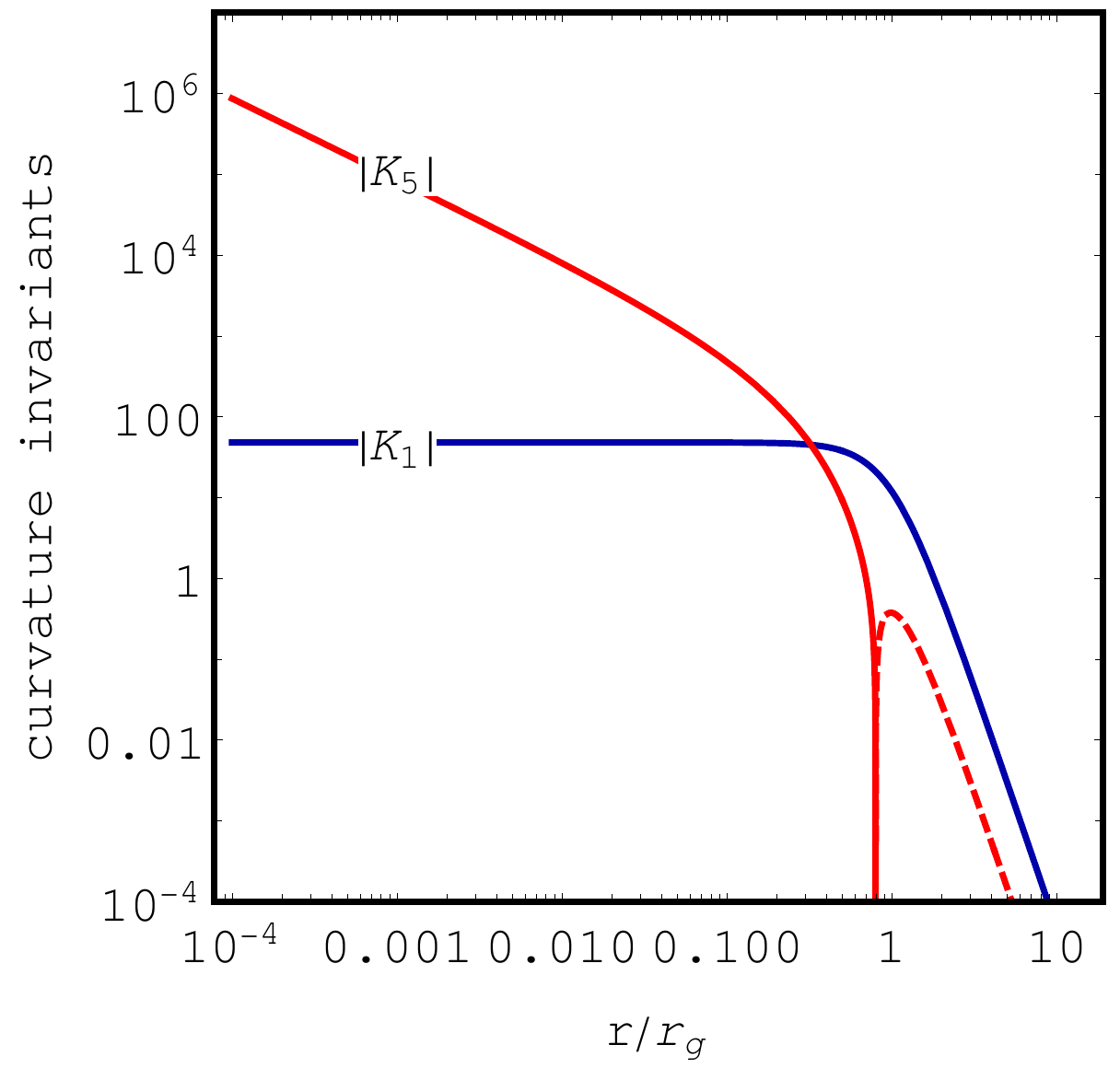}
	\caption{\label{fig:invariantRG-Schw} Characterization of the invariant-RG-improved Schwarzschild black hole. Left-hand panel: Metric function $A(r)$ for Schwarzschild (green dashed), near-critical $\ell_\text{NP}=3r_g$ (magenta), and diffent orders of magnitude $\ell_\text{NP}=10^{-n/2}r_g$ for $0\leqslant n\leqslant 4 \in \mathbb{N}$ (thin gray from right to left). $A(r)=0$ indicates the locations of event horizons. Right-hand panel: Two selected non-vanishing curvature invariants. (Dashing indicates negative values.) The Weyl invariants (e.g., $K_1$ in blue) are finite at $r=0$. The Ricci (and mixed) invariants (e.g., $K_5$ in red) remain divergent at $r=0$. All quantities are given in units of $r_g=G_0M$.
}
\end{figure}

To proceed with reconstructing the metric associated to a particular RG-improvement $G(\mathcal{K}_1(r))$, we make an ansatz for the most general static, spherically-symmetric spacetime
\begin{align}
\label{eq:sphSymm-ansatz}
	ds^2 = - A(r) dt^2 + \frac{1}{B(r)}dr^2 + r^2d\Omega\;,
\end{align}
where $A(r)$ and $B(r)$ remain to be determined\footnote{While a static and spherically-symmetric metric can in general have independent $tt$- and $rr$-components, a vanishing Ricci-tensor necessarily implies $g_{tt} = 1/g_{rr}$.}. The set of ZM-invariants $\widehat{\mathcal{K}}_i$ for this ansatz is given in App.~\ref{app:sph-symm-AB} where also some polynomial relations (so called syzygies) are listed. In general, these are of arbitrary degree and hence hard to establish. Indeed, we cannot exclude the potential existence of further ones. Such (polynomial or functional) relations might be connected to the question whether the resulting set of differential equations, i.e., $\widetilde{\mathcal{K}}_i = \widehat{\mathcal{K}}_i$, are overdetermined or not. 

Besides the physical assumption of dropping the requirement of Ricci-flatness (i.e., choosing to not enforce $\widetilde{\mathcal{K}}_{i\geqslant5} = 0$), we will focus the following discussion on the special case of $A\equiv B$. The exploration of $A\neq B$ may allow to determine other invariant-RG-improved spacetimes for which some of the Ricci-flat relations, i.e., $\widetilde{\mathcal{K}}_{i\geqslant5} = 0$, can be maintained. It remains an interesting question to investigate such potential $A\neq B$ solutions in the future.

For any given RG-improvement $G_0\rightarrow G(\mathcal{K}_1(r))$, the remaining free function $A(r)$ is determined by the RG-improved invariant $\widetilde{\mathcal{K}}_1$ (or equivalently $\widetilde{\mathcal{K}}_3$), i.e., by the differential equation
\begin{align}
\label{eq:AequalsB_ode1}
	\frac{\left(r^2 A''-2 r A'+2 A-2\right)^2}{3 r^4} = \widehat{\mathcal{K}}_1 \equiv \widetilde{\mathcal{K}}_1 =\frac{48G(\mathcal{K}_1(r))^2M^2}{r^6}\;.
\end{align}
Implementing the spherical limit of the RG-improvement reviewed in Sec.~\ref{sec:synopsis}, cf.~Eq.~\eqref{eq:scale-dependence} and~\eqref{eq:scale-identification} as well as~\cite{Bonanno:1998ye} for the original discussion in spherical symmetry, we obtain the resulting invariant-RG-improved spacetime specified by
\begin{align}
\label{eq:sol-invRG}
	A(r) = &
	1
	-\frac{2 r \ell}{\ell_\text{NP}^2} 
	\Bigg[
		+\sqrt{3} \pi
		\log(\ell^2-r \ell+r^2)
		-2 \log(\ell+r)	
		+2 \arctan\left(
			\frac{1-2 r/\ell}{\sqrt{3}}
		\right)
	\Bigg]
	\notag\\&   
   +\frac{4 r^2}{\ell_\text{NP}^2}\left[\log \left(\ell^3+r^3\right)-3 \log (r)\right]
   \quad\quad\quad\text{with}\quad\quad\quad
   \ell = \sqrt[3]{G_0M\, \ell_\text{NP}^2}\;.
\end{align}
Herein, we have fixed the constants of integration by demanding that $A(r)\rightarrow 1-2G_0M/r$ for large $r\gg M$. While this asymptotic behavior is not apparent in the exact form, it can be confirmed by expanding Eq.~\eqref{eq:sol-invRG} about $r=\infty$, cf.~also the left-hand panel in Fig.~\ref{fig:invariantRG-Schw}.

An expansion about $r=0$ confirms that $\mathcal{K}_{1,\,3}$ have finite limits, i.e., the divergence of the Kretschmann scalar in Schwarzschild spacetime is tamed by the RG-improvement. However, this does not hold for the Ricci and the mixed invariants $\mathcal{K}_{i\leqslant 5}$. Some of these are non-vanishing, i.e., $\mathcal{K}_{5-8,9,11,15,16}\neq 0$, since the invariant-RG-improved spacetime is no longer Ricci-flat. All of these diverge at $r=0$ like $\sim1/r^n$, with $n$ the number of Ricci tensors appearing in the respective curvature invariant. In particular, the Ricci scalar $R \equiv \mathcal{K}_5$ diverges like $1/r$, cf.~right-hand panel in Fig.~\ref{fig:invariantRG-Schw}. It remains an important open question whether this curvature singularity is related to geodesic incompleteness or not.
\\

We note that the above divergences of the invariant-RG-improved spacetime may be related to the special algebraic character of the classical Schwarzschild spacetime. On the one hand, the scale-identification, cf.~Eq.~\ref{eq:scale-identification} in the non-spinning $a\rightarrow0$ limit, relies on Ricci flatness. On the other hand, Ricci-flatness is broken in the invariant-RG-improved spacetime.
More generally, the classical curvature invariants, which determine the scale identification, are distinct from those of the RG-improved spacetime. As proposed in~\cite{Platania:2019kyx} in the context of metric-RG-improvement, iterative RG-improvement can lead to a converged spacetime in which, eventually, the RG-improved invariants agree with those that determine the scale identification. It thus seems promising to also iterate the proposed invariant-RG-improvement described above. Crucially, one should then also iteratively update the scale identification, cf.~Eq.~\ref{eq:scale-identification}, in order to include novel independent invariants which arise once Ricci-flatness is broken.

\subsection{Comments on invariant-RG-improved Kerr spacetime}

As demonstrated in the spherically symmetric case before, the straightforward part of invariant-RG-improvement is to RG-improve the curvature invariants. This also holds for Kerr spacetime. Given the only non-vanishing complex invariant $\mathbb{I}$, cf.~Eq.~\eqref{eq:II-Kerr} and App.~\ref{app:ZM-classical}, invariant-RG-improvement amounts to simply replacing $G_0\rightarrow G(|\mathbb{I}|)$ by a function of the invariant itself, cf.~Eq.~\ref{eq:scale-identification}. As for Schwarzschild spacetime, it remains non-trivial to reconstruct the respective metric. 
\\

On the one hand, just as for Schwarzschild spacetime, one has to \emph{decide} to break part of the algebraically special character. Without doing so, and since invariant-RG-improvement preserves all the relations among curvature invariants, a potential corresponding spacetime would remain Ricci-flat and of Petrov-type D. Uniqueness proofs for stationary solutions of the vacuum Einstein equations~\cite{Robinson:1975bv} directly imply that, without giving up either the Petrov type or Ricci flatness, no beyond-GR spacetime can be found.

On the other hand, one may decide to not break relations among the Weyl invariants. In addition, one can demand that the invariant-RG-improved spacetime retains a Killing-Yano tensor and an associated hidden constant of motion. The latter is a stronger condition than the former since it is known that a Killing-Yano tensor implies that the spacetime is Petrov-type D~\cite{1974Tenso..28..173C}, while there exist some Petrov-type D spacetimes without a Killing-Yano tensor~\cite{1980IJTP...19..675D}.

Stationarity, axisymmetry, and a hidden constant of motion imply that one can find coordinates in which the inverse metric can be expressed as~\cite{1979GReGr..10...79B}
\begin{align}
\label{eq:Benenti-Francaviglia-ansatz}
	g^{ab}\partial_{a}\partial_{b}  =\frac{1}{S_{r}+S_{\theta}}
	\Big[
		\left(G_{r}^{ij}+G_{\theta}^{ij}\right)\partial_{\sigma_i}\partial_{\sigma_j}
		+\Delta_{r}\partial_{r}^{2}
		+\Delta_{\theta}\partial_{\theta}^{2}
	\Big]\;,
\end{align}
where the coordinates $\sigma_i = (t,\,\phi)$ are the passive ones, associated with the Killing vectors, while $S_r(r)$, $G_r^{ij}(r)$, and $\Delta_r(r)$ as well as $S_\theta(\theta)$, $G_\theta^{ij}(\theta)$, and $\Delta_\theta(\theta)$ are functions of the active coordinates $r$ and $\theta$, respectively. Then, the associated Killing tensor $K^{ab}$ reads
\begin{align}
	K^{ab}\partial_{a}\partial_{b} = \frac{1}{S_{r}+S_{\theta}}
	\Big[
		\left(S_{r}G_{\theta}^{ij} - S_{\theta}G_{r}^{ij}\right)\partial_{\sigma_i}\partial_{\sigma_j}
		+S_{r}\Delta_{\theta}\partial_{\theta}^{2}
		-S_{\theta}\Delta_{r}\partial_{r}^{2}
	\Big]\;.
\end{align}
As one may check, cf.~\cite{1979GReGr..10...79B}, Boyer-Lindquist coordinates for the Kerr metric have this form. 
In principle, one can now calculate the ZM-invariants $\widehat{\mathcal{K}}_i$ for this ansatz and determine (for any explicit RG-improvement) whether the resulting differential equations for the two independent Weyl invariants have a solution.
\\

Even without invariant RG-improvement as a motivation, the introduced method may serve as a systematic construction principle for spinning spacetimes which (i) are regular at every spacetime point, (ii) have non-spherical horizons (i.e., introduce non-trivial angular-dependence in comparison to Kerr), and (iii) still exhibit a hidden constant of motion. The author is unaware of any examples of such metrics in the literature. See~\cite{Johannsen:2015pca, Papadopoulos:2018nvd} for examples that break regularity,~\cite{Eichhorn:2021iwq} for examples without (or at least without a known) Killing tensor, and~\cite{
Gambini:2013ooa,
Bambi:2013ufa,
Azreg-Ainou:2014pra,
Toshmatov:2014nya,
Ghosh:2014hea,
Fathi:2020agx,
Kumar:2020owy,
Mazza:2021rgq,
Junior:2021atr
} for examples without non-trivial angular dependence. The latter type of spacetimes can be generated via the Newman-Janis algorithm~\cite{Newman:1965tw,Drake:1998gf} from a non-spinning regular spacetime.
Invariant RG-improvement can be understood as a means to systematically determine whether the conditions (i)-(iii) can or cannot be simultaneously fulfilled. Overall, invariant RG-improvement of Kerr spacetime thus remains an exciting topic to continue in future work.

\section{Discussion}
\label{sec:discussion}

This work develops key insights regarding the general procedure of Renormalization-Group (RG) improvement of classical spacetimes. On the one hand, it establishes that RG improvement at the level of the metric is a coordinate-dependent procedure. On the other hand, it introduces a novel RG-improvement procedure at the level of curvature invariants which is manifestly coordinate independent. Invariant-RG improvement preserves all relations among curvature invariants and, in particular, the algebraic type of the classical spacetime. This also implies that in highly symmetric settings it may be necessary to make an additional physical choice about how to reduce the special algebraic type.

More generally, the presented results strongly suggest that any RG-improvement procedure which replaces couplings in non-scalar quantities is coordinate-dependent: Hence, it can be expected that RG-improvement at the level of the equations of motion is coordinate-dependent too. In contrast, while not explicitly shown here, there is no reason to expect that similar ambiguities arise for RG-improvement at the level of the action.
\\

The above statements are fully generic and apply to all cosmological as well as astrophysical spacetimes. More specifically, we explore spherically-symmetric and axially-symmetric black-hole spacetimes as explicit examples. In particular, we demonstrate the invariant RG-improvement procedure with the example of Schwarzschild spacetime. Here, we make the physical choice to give up Ricci flatness. Moreover, for simplicity, we restrict the RG-improved solution to be described by a single free metric function, i.e., $g_{rr} = 1/g_{tt}$. The resulting invariant RG-improved spacetime has regular Weyl invariants (which implies a regular Kretschmann scalar). However, no longer being Ricci-flat, all invariants involving the Ricci tensor become non-vanishing and divergent at $r=0$. For now, it remains an important open question whether the resulting spacetime is geodesically complete.
\\

Invariant RG-improvement opens up a way to incorporate scale-dependence of gravitational couplings into classical spacetimes in a coordinate-independent manner. It remains an exciting future avenue to explore the implications of such a procedure beyond the example of spherically-symmetric black holes.

\section*{Acknowledgements}
The author thanks Astrid Eichhorn for collaboration in a related project, motivating many considerations of this manuscript, for numerous insightful comments, and, in particular, for pivotal discussions regarding the noncommutative relation between coordinate transformations and metric RG-improvement. The author also thanks Alessia Platania for many helpful comments.

\paragraph{Funding information}
The author acknowledges funding by the Royal Society International Newton Fellowship NIF{\textbackslash}R1{\textbackslash}191008.

\begin{appendix}

\section{Zakhary-McIntosh (ZM) invariants}
\label{app:ZM-basis}

The set of Zakhary-McIntosh (ZM) invariants provides a polynomially complete basis of Riemann invariants $\mathcal{K} = \lbrace R,R_{\mu\nu},R_{\mu\nu\rho\sigma}R^{\mu\nu\rho\sigma},\dots\rbrace$~\cite{1991JMP....32.3135C, 1997GReGr..29..539Z, 2002nmgm.meet..831C}. It consists of four Weyl invariants, i.e.,
\begin{align}
	I_{1} &=C_{\mu\nu\rho\sigma}C^{\mu\nu\rho\sigma},
	\\
	I_{2} &=C_{\mu\nu\rho\sigma}\overline{C}^{\mu\nu\rho\sigma},
	\\
	I_{3} &=C_{\mu\nu}^{\phantom{\mu\nu}\rho\sigma}C_{\rho\sigma}^{\phantom{\rho\sigma}\alpha\beta }C_{\alpha\beta }^{\phantom{\alpha\beta }\mu\nu},
	\\
	I_{4} &=\overline{C}_{\mu\nu}^{\phantom{\mu\nu}\rho\sigma}C_{\rho\sigma}^{\phantom{\rho\sigma}\alpha\beta }C_{\alpha\beta }^{\phantom{\alpha\beta }\mu\nu},
\end{align}
four Ricci invariants, i.e.,
\begin{align}
	I_{5} &=R,
	\\
	I_{6} &=R_{\mu}^{\phantom{\mu}\nu}R_{\nu}^{\phantom{\nu}\mu},
	\\
	I_{7} &=R_{\mu}^{\phantom{\mu}\nu}R_{\nu}^{\phantom{\nu}\rho}R_{\rho}^{\phantom{\rho}\mu},
	\\
	I_{8} &=R_{\mu}^{\phantom{\mu}\nu}R_{\nu}^{\phantom{\nu}\rho}R_{\rho}^{\phantom{\rho}\sigma}R_{\sigma}^{\phantom{\sigma}\mu},
\end{align}
and nine mixed Ricci-Weyl invariants, i.e.,
\begin{align}
	I_{9} &=R^{\mu\nu}R^{\rho\sigma}C_{\mu\nu\rho\sigma},
	\\
	\label{eq:I10}
	I_{10} &=R^{\mu\nu}R^{\rho\sigma}\overline{C}_{\mu\nu\rho\sigma},
	\\
	I_{11} &=R^{\nu\rho}R_{\gamma\delta }\left(
		C_{\mu\nu\rho\sigma}C^{\mu \gamma\delta \sigma} 
		- \overline{C}_{\mu\nu\rho\sigma}\overline{C}^{\mu\gamma\delta \sigma}
	\right),
	\\
	\label{eq:I12}
	I_{12} &= 2 R^{\nu\rho}R_{\gamma\delta }C_{\mu\nu\rho\sigma}\overline{C}^{\mu\gamma\delta\sigma},
	\\
	I_{13} &=R_{\mu}^{\phantom{\mu}\gamma}R_{\gamma}^{\phantom{\gamma}\rho}R_{\nu}^{\phantom{\nu}\delta}R_{\delta}^{\phantom{\delta}\sigma}C^{\mu\nu}_{\phantom{\mu\nu}\rho\sigma},
	\\
	I_{14} &=R_{\mu}^{\phantom{\mu}\gamma}R_{\gamma}^{\phantom{\gamma}\rho}R_{\nu}^{\phantom{\nu}\delta}R_{\delta}^{\phantom{\delta}\sigma}\overline{C}^{\mu\nu}_{\phantom{\mu\nu}\rho\sigma},
	\\
	I_{15} &=\frac{1}{16}R^{\nu\rho}R_{\gamma\delta}\left(
		C_{\mu\nu\rho\sigma}C^{\mu\gamma\delta\sigma} 
		+ \overline{C}_{\mu\nu\rho\sigma}\overline{C}^{\mu\gamma\delta\sigma}
	\right),
	\\
	\label{eq:ZM-I16}
	I_{16} &=\frac{1}{32}R^{\rho\sigma}R^{\gamma\delta }C^{\mu\kappa\lambda\nu}\left(
		C_{\mu\rho\sigma\nu}C_{\kappa\gamma\delta\lambda} 
		+ \overline{C}_{\mu\rho\sigma\nu}\overline{C}_{\kappa\gamma\delta\lambda}
	\right),
	\\
	I_{17} &=\frac{1}{32}R^{\rho\sigma}R^{\gamma\delta }\overline{C}^{\mu \kappa\lambda \nu}\left(
		C_{\mu\rho\sigma\nu}C_{\kappa\gamma\delta\lambda} 
		+ \overline{C}_{\mu\rho\sigma\nu}\overline{C}_{\kappa\gamma\delta\lambda}
	\right)\;,
\end{align}
and suffices to distinguish all Petrov~\cite{1954UZKGU.114...55P} and Segre~\cite{Stephani:2003tm} types. Herein, $R$, $C$ and $\overline{C}$ denote the Ricci, the Weyl, and the (left-)dual Weyl tensor $\overline{C}_{\mu\nu\rho\sigma} = 1/2\,\epsilon_{\mu\nu \kappa\lambda }C^{\kappa\lambda }_{\phantom{\kappa\lambda }\rho\sigma}$, respectively.
Alternatively, the set may be expressed in terms of six complex invariants, i.e.,
\begin{align}
	\mathbb{I} &= I_1 + i I_2\;,
	\\
	\mathbb{J} &= I_3 + i I_4\;,
	\\
	\mathbb{K} &= I_9 + i I_{10}\;,
	\\
	\mathbb{L} &= I_{11} + i I_{12}\;,
	\\
	\mathbb{M} &= I_{13} + i I_{14}\;,
	\\
	\mathbb{M}_2 &= I_{16} + i I_{17}\;,
\end{align}
and five remaining real invariants, i.e.,
\begin{align}
	I_5\;,\quad
	I_6\;,\quad
	I_7\;,\quad
	I_8\;,\quad
	I_{15}\equiv\mathbb{M}_1\;.
\end{align}
For the examples we have looked at in this work, the algebraic complexity of the expressions is most reduced in the latter form. Explicit expressions may be obtained from a specified metric using computer algebra toolkits such as GRTensor~\cite{1994grra.conf..317L} or xAct~\cite{Martin-Garcia:2007bqa, MartinGarcia:2008qz}. In the following, we will characterize spacetimes by these six complex and five real invariants. Equivalently, we provide the 17 real ZM invariants in ancillary files~\cite{anc}.

\section{ZM invariants: classical black-hole spacetimes}
\label{app:ZM-classical}

The independent invariants of Schwarzschild and Kerr spacetime are, of course, long known, see e.g.~\cite{Overduin:2020aiq}. For completeness, we provide them in the following. We present them as limits of the Kerr-Newman spacetime which gives us the opportunity to relate the invariants to the complexification procedure known as Newman-Janis algorithm \cite{Newman:1965tw}. A somewhat similar complexification also relates the metric-RG-improved Schwarzschild and Kerr spacetime -- if both are obtained in horizon-penetrating coordinates, i.e., EF and ingoing Kerr coordinates, respectively, cf.~Sec.~\ref{app:ZM-metric-RG}.
\\

Kerr-Newman spacetime with mass $M$ as well as normalized spin $a$ and charge $q$ (using natural units in which $c\!\equiv\!\hbar\!\equiv\!4\pi\epsilon_0\!\equiv\! 1$) is characterized by two non-trivial invariants, i.e.,
\begin{align}
	\mathfrak{C}^6\!\equiv\!
	\left[\frac{\mathbb{I}}{2^4\,3}\right]^3 \!=\!
	\left[\frac{\mathbb{J}}{2^5\,3}\right]^2 \!=\!
	\left[\frac{
		G_0q^2-G_0M (r+i a \chi )
	}{
		(r-i a \chi )^3 (r+i a \chi)
	}\right]^6 & \!=\!
	\left[\frac{\mathbb{K}}{2^2 I_6}\right]^6 \!=\! 
	\left[\frac{\mathbb{L}}{2^2 \mathbb{K}}\right]^6 \!=\! 
	\left[\frac{2^6\,3\,\mathbb{M}_2^2}{\mathbb{I}\overline{\mathbb{K}}}\right]^3\;,
	\\
	\mathfrak{R}^4\!\equiv\!
	\left[\frac{
		G_0q^2
	}{
		(r-i a \chi )^2 (r+i a \chi)^2
	}\right]^4 &\!=\! 
	\frac{I_8}{2^2} \!=\!
	\left[\frac{I_6}{2^2}\right]^2 \!=\!
	\left[\frac{\mathbb{K}\,\overline{\mathbb{K}}}{2^6\mathbb{M}_1}\right]^2	
	\;,
\end{align}
the first of which is complex and the second real. The remaining invariants vanish, i.e., $I_5 = I_7 = \mathbb{M}_1 = \mathbb{M} = 0$. We present the above invariants using the active coordinates $(r,\theta)$ of, for instance, ingoing Kerr (or equivalently Boyer-Lindquist) coordinates, introducing the `rational polynomial' coordinate $\chi\equiv\cos(\theta)$ as a shorthand. Also, we denote them in a suggestive form with Weyl-invariants on the left and Ricci (and mixed) invariants on the right of the explicit expression. 

One can confirm successively that the relations imply that only the Weyl-invariants (on the left) remain non-vanishing in the $q\rightarrow0$ limit. As demanded by consistency, the $q\rightarrow0$ limit recovers Kerr spacetime, the $a\rightarrow0$ limit recovers Reissner-Nordstr\"om spacetime, and the combined $q\rightarrow0$ and $a\rightarrow0$ limit recovers Schwarzschild spacetime, as expected. Vice versa, one can obtain the spinning solutions from the non-spinning ones by a particular complexification, cf.~\cite{Newman:1965tw}, although the choice of when to use $r\rightarrow r - ia\chi$ and when to use $r\rightarrow r + ia\chi$ remains unexplained, cf.~also~\cite{Drake:1998gf}

\section{ZM invariants: static spherically-symmetric spacetimes}
\label{app:sph-symm-AB}

For a general static and spherically-symmetric spacetime (no longer necessarily a vacuum solution of GR), there is more than one independent ZM invariant. Primes denote derivatives with respect to $r$ in the following. The Weyl invariants are determined by
\begin{align}
	\left(\frac{\mathbb{I}}{48}\right)^3 = \left(\frac{\mathbb{J}}{96}\right)^2 &=
	\frac{\left(A^2 \left(2 r B'-4 B+4\right)+B r^2 \left(A'\right)^2-A r \left(2 B r A''+A' \left(r
   B'-2 B\right)\right)\right)^6}{(2\sqrt{6}\,r\,A)^{12}}\;.
\end{align}
The other mixed invariants are given by
\begin{align}
	\frac{\mathbb{K}^2}{\mathbb{I}} =& 
	\frac{\left(A^2 \left(2 r B'-4 B+4\right)-B r^2 \left(A'\right)^2+A r \left(2 B r A''+A' \left(r B'-2 B\right)\right)\right)^2}{3\,(2r\,A)^8}\times
	\notag\\&
	\times\left(
		2 A^2 \left(r B'+2 B-2\right)
		+B r^2 \left(A'\right)^2
		-A r \left(r A' B'+2 B\left(r A''+A'\right)\right)
	\right)^2	
	\;,
	\\[0.5em]
	\frac{\mathbb{L}}{\mathbb{I}} =
	2^4\frac{\mathbb{M}_1}{\mathbb{I}} =&
	\frac{1}{2^4\;3}\Bigg[
		\frac{4 B^2 r^2 \left(A''\right)^2+\left(A'\right)^2 \left(r^2 \left(B'\right)^2+2 B (5 B-4)\right)+4 B r^2 A' A''B'}{A^2 r^2}
		\notag\\&\quad\quad\quad
		+\frac{4 \left((2-3 B) A' B'-4 (B-1) B A''\right)}{A r^2}
		-\frac{2 B \left(A'\right)^2 \left(2 B A''+A' B'\right)}{A^3}
		\notag\\&\quad\quad\quad
		+\frac{2 \left(r^2 \left(B'\right)^2+8 (B-1)^2\right)}{r^4}
		+\frac{B^2 \left(A'\right)^4}{A^4}
	\Bigg]
	\;,
	\\[0.5em]
	\frac{\mathbb{M}}{\mathbb{K}} =& \frac{\left(B A'-A B'\right)^2}{4 A^2 r^2}
	\;,
	\\[0.5em]
	\frac{{\mathbb{M}_2}^2}{\mathbb{I}^3} =& 
	\frac{\left(A^2 \left(r B'-4 B+4\right)-B r^2 \left(A'\right)^2+A r \left(2 B r A''+A' \left(r B'-B\right)\right)\right)^2}{3^3(2^3r\,A)^8}	\times
	\notag\\&
	\left(A^2 \left(r B'+4 B-4\right)+B r^2 \left(A'\right)^2-A r \left(2 B r A''+A' \left(r
   B'+B\right)\right)\right)^2
	\;.
\end{align}
The Ricci invariants are given by
\begin{align}
	I_{4+n} = &
	\left(\frac{-1}{4r^2 A(r)^2}\right)^n
	\Bigg[
		2^{n+1} A(r)^n \left(
			r B(r)A'(r)+A(r) \left(
				r B'(r)+2 B(r)-2
			\right)
		\right)^n
	\\\notag &
		+r^n \left(
			A(r) \left(
				r A'(r)+4 A(r)
			\right)
			B'(r)-r B(r) \left(
				A'(r)^2-2 A(r) A''(r)
			\right)
		\right)^n
	\\\notag &
		+r^n \left(
			(-1)^n \left(
				r B(r)A'(r)^2-A(r) \left(
					2 r B(r) A''(r)+A'(r) \left(
						r B'(r)+4B(r)
					\right)
				\right)
			\right)
		\right)^n
	\Bigg]\;,
\end{align}
for $1\leqslant n\leqslant 4$. While we have not attempted to analytically prove it, we conjecture that this formula holds for arbitrary numbers of contractions of Ricci tensors.

\section{ZM invariants: distinct metric-RG-improved spacetimes}
\label{app:ZM-metric-RG}

Here, we provide the ZM invariants for metric-RG-improved Schwarzschild and Kerr spacetime for several different sets of coordinates of the classical spacetime. In some cases, we only present those invariants which are required to obtain the polynomial relations used in Sec.~\ref{sec:metric-RG} to prove that the spacetimes are inequivalent and therefore that metric-RG improvement is a coordinate-dependent procedure.

\subsection{Spherical symmetry}
\label{app:ZM_metric-RG_spherical}

\subsubsection{Metric-RG-improved spacetime rooted in EF coordinates}
\label{app:ZM_metric-RG_EF}

Having performed metric RG-improvement of Schwarzschild spacetime in EF or Schwarzschild coordinates, the resulting metric-RG-improved spacetime is characterized in terms of independent ZM invariants by
\begin{align}
	\mathbb{I}^3 = 12\mathbb{J}^2 &= 
	\left[\frac{2M \left(r \left(r G''-4 G'\right)+6 G\right)}{\sqrt{3}r^{3}}\right]^6\;,
	\notag\\
	\frac{\mathbb{K}^2}{\mathbb{I}} = 
	\frac{3\,\mathbb{L}^2}{\mathbb{I}^2} = 
	\frac{3\times 2^8\,\mathbb{M}_1^2}{\mathbb{I}^2} = 
	\frac{3^2\times 2^{10}\,\mathbb{M}_2^2}{\mathbb{I}^3} &=
	\frac{1}{3}\left[\frac{M \left(r G''-2 G'\right)}{r^2}\right]^4\;,
	\notag\\
	\mathbb{M} &=0\;,
	\notag\\
	\label{eq:invariants_metric-RG-Schw_EF-coords}
	\mathcal{K}_{4+n} &= \frac{2 M^n\left(r^n \left(G''\right)^n+2^n \left(G'\right)^n\right)}{r^{2 n}}\;,
\end{align}
where $1\leqslant n\leqslant 4$ in the last line and primes denote derivatives of the scale-dependent Newton coupling $G(r)$ with respect to $r$. The four invariants in Eq.~\eqref{eq:invariants_metric-RG-Schw_EF-coords} are related by the following two syzygies
\begin{align}
	\mathfrak{K}_1 &= \frac{1}{8}\left(\mathcal{K}_5^2 - 2\,\mathcal{K}_6\right)^2 - \left(\mathcal{K}_6^2 - 2\,\mathcal{K}_8\right)
	= 0\;,
	\notag\\
	\mathfrak{K}_2 &= \frac{1}{8}\mathcal{K}_5\left(\mathcal{K}_5^2 - 6\,\mathcal{K}_6\right)^2 +\mathcal{K}_7 
	= 0\;.
	\label{eq:syzygies}
\end{align}

Since the coordinate transformation is the exception to the rule and commutes with metric RG-improvement, cf.~Sec.~\ref{sec:metric-RG}, one may check that the above invariants also characterize metric RG-improvement of Schwarzschild spacetime in Schwarzschild coordinates.

\subsubsection{Metric-RG-improved spacetime rooted in modified EF coordinates}
\label{app:ZM_metric-RG_mod-EF}

Performing metric RG-improvement of Schwarzschild spacetime in \emph{modified} EF coordinates (with integer parameter $n$, cf.~Eq.~\eqref{eq:lineElem_EF-mod}), the Ricci invariants of the resulting metric-RG-improved spacetime read
\begin{align}
	\mathcal{K}_{4+m} =& 
	\frac{(G M)^{2 m n}}{G^{4 m} (nM)^{2 m} r^{3 m n}}\Bigg[
		(nr^2GG'')^m (G M)^{m n}
	\\&\notag
		+2 (-1)^m r^m\Big(
			2 G (G M)^n \left((n-1) r G'' - G'\right)
			\\\notag&\quad\quad\quad\quad\quad\quad
			-(n-1) r^n \left(G r G''+G' \left(G-n r G'+2 G n\right)\right)
		\Big)^m
	\\\notag&
		+\Big(
			nr^2GG'' (G M)^n
			+2r(n-1) \left(r^n-2 (G M)^n\right)\left(
				G \left(r G''+G'\right)
				-r G'^2
			\right)
		\Big)^m
	\Bigg]\;.
\end{align}
for $1\leqslant m\leqslant 4$. We refrain from writing out the remaining ZM invariants. Note, however, that
\begin{align}
	\mathfrak{K}_2 =&
	\frac{3 (n-1)^2 r^{3-9 n} (G M)^{6 n}}
	{(n\,G^2M)^6}
	\times
	\left[r^n-2 (G M)^n\right]^2
	\times
	\left[r \left(G'\right)^2-G \left(r G''+G'\right)\right]^2
	\times
	\\\notag\times&
	\left[
		(n-1) r G'^2 \left(2 (G M)^n+(n-1) r^n\right)
		+G n \left(r G''(GM)^n-2 G'\left((G M)^n+(n-1) r^n\right)\right)
	\right]\;.
\end{align}
which is non-vanishing for $n\neq 1$. (The same holds for $\mathfrak{K}_1$ which we do not write out here.)

\subsection{Axial symmetry}
\label{app:ZM_metric-RG_axial}

\subsubsection{Metric-RG-improved spacetime rooted in Kerr coordinates}
\label{app:ZM_metric-RG_ingoingKerr}

For metric RG-improvement of Kerr spacetime rooted in horizon-penetrating ingoing Kerr coordinates, the resulting metric-RG-improved spacetime is characterized in terms of independent ZM invariants by
\begin{align}
	\mathbb{I}^3 &= 12\mathbb{J}^2 = 
	\frac{\left(2M(r-i a \chi ) \left(r (r-i a \chi )G'' -2(2 r+i a \chi )G'\right)+6(r+i a \chi )G\right)^6}{\left(\sqrt{3}(r-i a \chi )^{3} (r+i a \chi )\right)^3}\;,
	\notag\\
	\frac{\mathbb{K}^2}{\mathbb{I}} &= 
	\frac{3\,\mathbb{L}^2}{\mathbb{I}^2} = 
	\frac{3\; 2^8\,\mathbb{M}_1^2}{\overline{\mathbb{I}}\;\mathbb{I}} = 
	\frac{3^2\; 2^{10}\,\overline{\mathbb{M}}_2^2}{\overline{\mathbb{I}}^2\;\mathbb{I}} =
	\frac{1}{3}\left[\frac{
		M\left(r\left(r^2+a^2 \chi ^2\right)G''-2(r^2-a^2 \chi^2 )G'\right)
	}{
		\left(r^2+a^2 \chi ^2\right)^2
	}\right]^4\;,
	\notag\\
	\mathbb{M} &=0\;,
	\notag\\
	\label{eq:invariants_metric-RG-Kerr-Kerr-coords}
	\mathcal{K}_{4+n} &= 
	\frac{
		2 M^n\left(
			\left(r \left(r^2+a^2 \chi ^2\right) G''+2 a^2 \chi ^2 G'\right)^n
			+\left(2 r^2 G'\right)^n
		\right)
	}{
		\left(r^2+a^2 \chi^2\right)^{2 n}
	}\;,
\end{align}
where $1\leqslant n\leqslant 4$ in the last line, we introduced the shorthand $\chi = \cos(\theta)$, and primes denote derivatives of the scale-dependent Newton coupling $G(r,\chi)$ with respect to $r$. As for RG-improved Schwarzschild spacetime rooted in horizon-penetrating ingoing EF coordinates, the four invariants in Eq.~\eqref{eq:invariants_metric-RG-Schw_EF-coords} are related by the syzygies in Eq.~\eqref{eq:syzygies}.

We also highlight that the invariants of metric-RG-improved Kerr spacetime rooted in Kerr coordinates, cf.~Eq.~\eqref{eq:invariants_metric-RG-Kerr-Kerr-coords}, follow from those of metric-RG-improved Schwarzschild spacetime rooted in EF coordinates, cf.~Eq.~\eqref{eq:invariants_metric-RG-Schw_EF-coords}, by a complexification prescription. However, it is a non-trivial and apparently random choice which occurrences of $r$ to replace by $r+ia\chi$, which by $r-ia\chi$, and which not at all. This reflects the fact that the spinning metric RG-improved spacetime (in horizon-penetrating coordinates) does no longer follow from an NJ-algorithm procedure from the spherically-symmetric one.

\subsubsection{Metric-RG-improved spacetime rooted in BL coordinates}
\label{app:ZM_metric-RG_BL}

For metric RG-improvement of Kerr spacetime rooted in BL coordinates, the curvature invariants depend on $\theta$-derivatives of the RG-improved Newton coupling. This leads to a significant increase in the complexity of the expressions. Hence, we only list the four Ricci invariants and restrict to the limit $\cos(\theta)\equiv\chi\rightarrow 1$, i.e., $\theta\rightarrow n\,\pi$ with $n\in\mathbb{N}$. Denoting $n_r$ $r$- and $n_\chi$ $\chi$-derivatives of $G(r,\chi)$ with superscripts $^{(n_r,n_\chi)}$, we find
\begin{align}
	\mathcal{K}_5 =&
	\frac{2 M r G^{(2,0)}+4 M G^{(1,0)}}{a^2+r^2}\;,
	\\
	\mathcal{K}_6 =&
	\frac{8 M^2 r^2 {G^{(0,1)}}^2}{\left(a^2+r^2\right)^2 \left(a^2-2 G M r+r^2\right)^2}
	+\frac{8 a^2 M^2 r G^{(1,0)} G^{(2,0)}}{\left(a^2+r^2\right)^3}
	+\frac{2 M^2 r^2 {G^{(2,0)}}^2}{\left(a^2+r^2\right)^2}
	\notag\\&
	+\frac{8 M^2{G^{(1,0)}}^2 \left(a^4+r^4\right)}{\left(a^2+r^2\right)^4}\;,
	\\
	\mathcal{K}_7 =&
	\frac{48 a^2 M^3 r^2 {G^{(0,1)}}^2 G^{(1,0)}}{\left(a^2+r^2\right)^4 \left(a^2-2 G M r+r^2\right)^2}
	+\frac{24 M^3 r^3 {G^{(0,1)}}^2 G^{(2,0)}}{\left(a^2+r^2\right)^3 \left(a^2-2 G M r+r^2\right)^2}
	\notag\\&
	+\frac{2 M^3 r^3 {G^{(2,0)}}^3}{\left(a^2+r^2\right)^3}
	+\frac{12 a^2 M^3r^2 G^{(1,0)} {G^{(2,0)}}^2}{\left(a^2+r^2\right)^4}
	\notag\\&
	+\frac{24 a^4 M^3 r {G^{(1,0)}}^2 G^{(2,0)}}{\left(a^2+r^2\right)^5}
	+\frac{16 M^3 {G^{(1,0)}}^3 \left(a^4-a^2r^2+r^4\right)}{\left(a^2+r^2\right)^5}\;,
	\\
	\mathcal{K}_8 =&
	\frac{32 M^4 r^4 {G^{(0,1)}}^4}{\left(a^2+r^2\right)^4 \left(a^2-2 G M r+r^2\right)^4}
	+\frac{48 M^4 r^4 {G^{(0,1)}}^2 {G^{(2,0)}}^2}{\left(a^2+r^2\right)^4 \left(a^2-2 G M r+r^2\right)^2}
	\notag\\&
	+\frac{192 a^2 M^4 r^3{G^{(0,1)}}^2 G^{(1,0)} G^{(2,0)}}{\left(a^2+r^2\right)^5 \left(a^2-2 G M r+r^2\right)^2}
	+\frac{192 a^4 M^4 r^2 {G^{(0,1)}}^2{G^{(1,0)}}^2}{\left(a^2+r^2\right)^6 \left(a^2-2 G M r+r^2\right)^2}
	\notag\\&
	+\frac{2 M^4 r^4 {G^{(2,0)}}^4}{\left(a^2+r^2\right)^4}
	+\frac{16 a^2 M^4 r^3 G^{(1,0)}{G^{(2,0)}}^3}{\left(a^2+r^2\right)^5}
	+\frac{32 M^4 {G^{(1,0)}}^4 \left(a^8+r^8\right)}{\left(a^2+r^2\right)^8}
	\notag\\&
	+\frac{64 a^6 M^4 r {G^{(1,0)}}^3 G^{(2,0)}}{\left(a^2+r^2\right)^7}
	+\frac{48 a^4 M^4 r^2 {G^{(1,0)}}^2 {G^{(2,0)}}^2}{\left(a^2+r^2\right)^6}\;.
\end{align}
The full expressions (without taking the $\chi\rightarrow 1$ limit) can be found in the ancillary files. In contrast to the horizon-penetrating ingoing-Kerr coordinate case, cf.~App.~\ref{app:ZM_metric-RG_ingoingKerr}, derivatives with respect to the angular coordinate $\chi$ do \emph{not} cancel out in the final expressions.

The syzygies that vanish for metric-RG-improved Kerr spactime rooted in Kerr coordinates, cf.~Eq.~\ref{eq:syzygies}, no longer vanish, i.e., (in the limit of $\chi\rightarrow1$)
\begin{align}
	\mathfrak{K}_1 =&
	\frac{32 M^4 r^4 {G^{(0,1)}}^4}{\left(a^2+r^2\right)^4 \left(a^2-2 G M r+r^2\right)^4}
	+\frac{64 M^4 r^4 {G^{(0,1)}}^2 {G^{(2,0)}}^2}{\left(a^2+r^2\right)^4 \left(a^2-2 G M r+r^2\right)^2}
	\\\notag&
	-\frac{64 M^4 r^3 {G^{(0,1)}}^2G^{(1,0)} G^{(2,0)} \left(r^2-4 a^2\right)}{\left(a^2+r^2\right)^5 \left(a^2-2 G M r+r^2\right)^2}
	-\frac{128 M^4 r^2 {G^{(0,1)}}^2 {G^{(1,0)}}^2 \left(-2 a^4+a^2 r^2+r^4\right)}{\left(a^2+r^2\right)^6 \left(a^2-2
   G M r+r^2\right)^2}
	\;,
	\\
	\mathfrak{K}_2 =&
	\frac{24 M^3 r^2 {G^{(0,1)}}^2 G^{(1,0)} (a-r) (a+r)}{\left(a^2+r^2\right)^4 \left(a^2-2 G M r+r^2\right)^2}
	+\frac{12 M^3 r^3 {G^{(0,1)}}^2 G^{(2,0)}}{\left(a^2+r^2\right)^3 \left(a^2-2 G M r+r^2\right)^2}
	\;,
\end{align}
which establishes the inequivalence of the two metric-RG-improved spacetimes. Finally, note that some of the invariants harbor curvature singularities at $\left(a^2-2 G M r+r^2\right)=0$.

\end{appendix}

\bibliography{References}

\nolinenumbers

\end{document}